\documentstyle[12pt,epsfig]{article}
\def\lsim{\mathrel{\rlap{
\lower4pt\hbox{\hskip-3pt$\sim$}}
    \raise1pt\hbox{$<$}}}     %less than approx. symbol
\def\gsim{\mathrel{\rlap{
\lower4pt\hbox{\hskip-3pt$\sim$}}
    \raise1pt\hbox{$>$}}}     %greater than or approx. symbol

\begin{document}

\begin{center}
{\large {\bf
Observation of New Resonance Structure in the Invariant Mass Spectrum of Two
$\gamma$-Quanta in dC-Interactions at Momentum 2.75 GeV/c per Nucleon
}}

\vspace*{5mm}
Kh.U.~Abraamyan$^{a)}$, A.N.~Sissakian$^{b)}$, and A.S.~Sorin$^{b)}$\\

\vspace*{3mm} {\it
$a)$ VBLHE JINR, 141980 Dubna, Moscow region, Russia\\
abraam@sunhe.jinr.ru\\
$b)$ BLTP  JINR, 141980 Dubna, Moscow region, Russia\\
sisakian@jinr.ru\\
sorin@theor.jinr.ru
}
\end{center}

\vspace*{5mm}

{\small { \centerline{\bf Abstract}

A new resonance structure at $M= 355 \pm 6 \pm 9$ MeV is observed in the
invariant mass spectrum of two $\gamma$-quanta
produced in the reaction $d + C\to\gamma + \gamma +x$ at momentum 2.75 GeV/c per nucleon.
Preliminary estimates of its width and cross section are $\Gamma = 41 \pm 12$ MeV and
$\sigma_{\gamma\gamma}\sim 0.6 ~ \mu b$. The collected statistics is
$2680 \pm 310$ events of $1.5\cdot 10^6$ triggered interactions of a total number
$3\cdot 10^{12}$ dC-interactions.
}}\\[5mm]

\vspace*{3mm}

\section{Introduction}

In connection with the recently developed physics program [1] on the search for a manifestation
of a mixed phase of strongly interacting QCD matter in nucleus-nucleus collisions at the
Nuclotron, the JINR initiated investigations aimed at finding and studying possible nontrivial
structures in the spectrum of invariant effective masses of two photons in the region starting with
the two-pion threshold to the $\rho$-meson mass. This program can be considered as a pilot one
for further detailed studies at SIS-100/300 (GSI) and as a part of the international research
program on studying the energy dependence of the properties of matter produced in collisions of
relativistic nuclei. In this case, it is extremely important to study the evolution of two-photon
spectra not only with respect to energy of colliding nuclei, but also to their atomic number.

It is to be noted that starting in 1997, data acquisition on production of neutral mesons and
$\gamma$-quanta in proton-nucleus and nucleus-nucleus interactions for light nuclei has been
carried out with internal beams of the Nuclotron in order to elucidate their production mechanism [2,3].
Experiments were conducted with internal proton beams at momentum 5.5 GeV/c incident on a carbon
target and with $^2H$, $^4He$ beams and internal C-, Al-, Cu-, W-,
and Au-targets at momenta from 1.7 to 3.8 GeV/c per nucleon. In the light of the ideas recently
arisen, analysis of these experimental data may prove to be the first important step in realizing
the above program, which is the main impetus of this paper.

For the first analysis there were selected data on dC-interactions at energy
$T_d = 2.0$ GeV/nucleon (momentum 2.75 GeV/c per nucleon) - data with maximum statistics.

Analysis of the data led to an intriguing
result - a peak was detected which however has not been observed in analogous analysis of other
experiments [2-5]. One of the possible explanations is the available statistics in the above-said
experiment which essentially exceeds that in other experiments analysed. The requirements imposed
on statistics for resolution of signals over the background were discussed in [3].
\begin{figure}[t]
 \centerline{ \includegraphics[width=7cm]{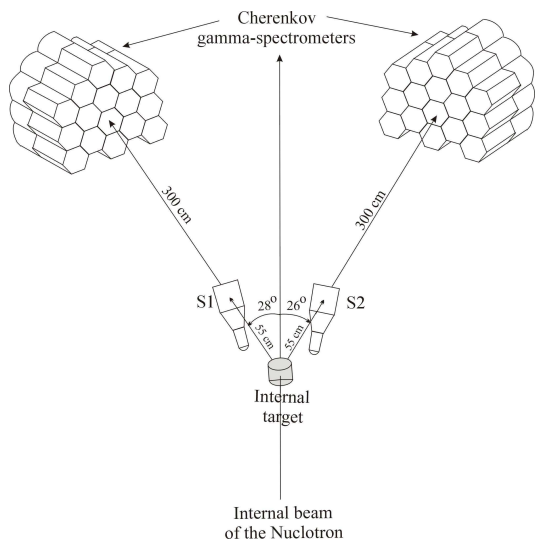}}
  \caption{The scheme of the experimental instrumentation }
\end{figure}

In this paper, the results are presented of the analysis of experimental data on production of
correlated pairs of $\gamma$-quanta in the reaction
$$d+C\to\gamma+\gamma +x, \eqno(1)$$
at momentum of incident deuterons 2.75 GeV/c per nucleon which were obtained with the 90-channel
Cherenkov gamma-spectrometer at the JINR Laboratory of High Energies (setup Photon-2) [2,3,6]. Due to
increased interest in the region of effective invariant masses 280-750 MeV [1], the data were analysed
under the conditions with a minimal background in the interval 300-400 MeV.

The paper is organized as follows. Section 2 is a brief description of the experiment
and experimental
setup. The main spectra of invariant effective masses of pairs of $\gamma$-quanta in reaction (1)
are discussed in Section 3.
Section 4 is the analysis of systematic errors in measurements of $\gamma$-quanta energies
and uncertainties coming from
describing the background. To elucidate the nature of the peak observed, the dependence of
its position and width on the opening angle of $\gamma\gamma$ pairs, energy of individual
$\gamma$-quanta, and the summed energy of the two $\gamma$-quanta is analysed in Section 5.
The estimates of the width and production cross section of the observed resonance structure are
given in Section 6. The main inferences of the paper are presented in the Conclusion.

\section{Experiment}

The experiment was carried out with an internal beam of deuterons with momentum 2.75 GeV/c per nucleon
and intensity $\sim 10^9$ particle/cycle of the Nuclotron. The experimental instrumentation allowed one
to measure both energies and the direction of emission of $\gamma$-quanta produced in reaction (1).
The experimental instrumentation is schematically represented in Fig.1.
The setup includes 32 $\gamma$-spectrometers of lead glass and scintillation counters $S_1$ and $S_2$ of
15 x 15 cm [6,7].

The modules of the $\gamma$-spectrometer are assembled into two arms with 16 units in each arm.
The detectors in each arm are divided into two groups with 8 units in each group. The output signals
in each group are summed up linearly and after discrimination by amplitude are used in fast triggering.
In this experiment the discriminator thresholds were at the level of 0.4 GeV. The setup was triggered
by the coincidence of signals from two or more groups of detectors in different arms.

\begin{figure}[t]
 \centerline{\includegraphics[width=7.5cm]{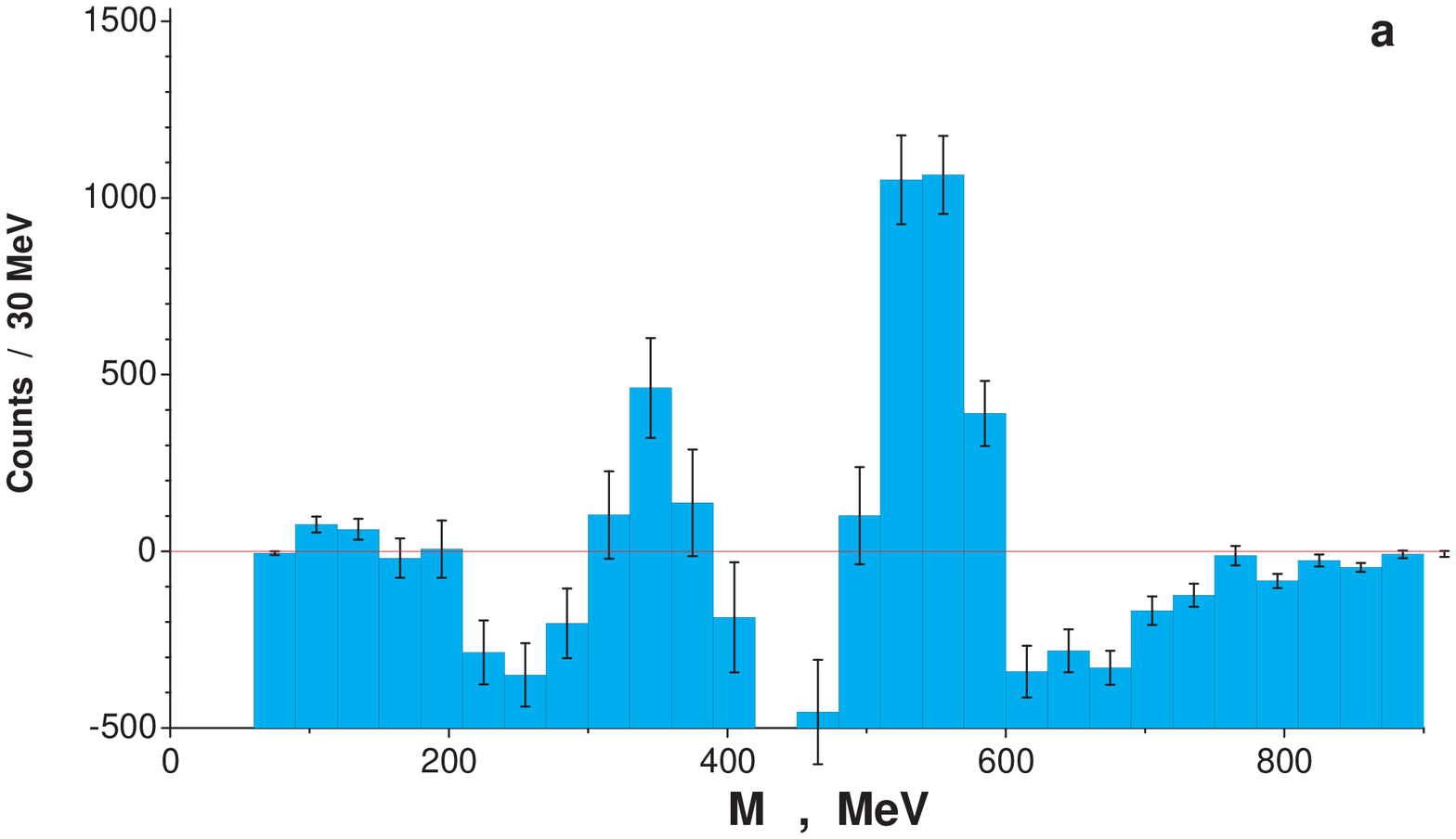}
 \includegraphics[width=7.5cm]{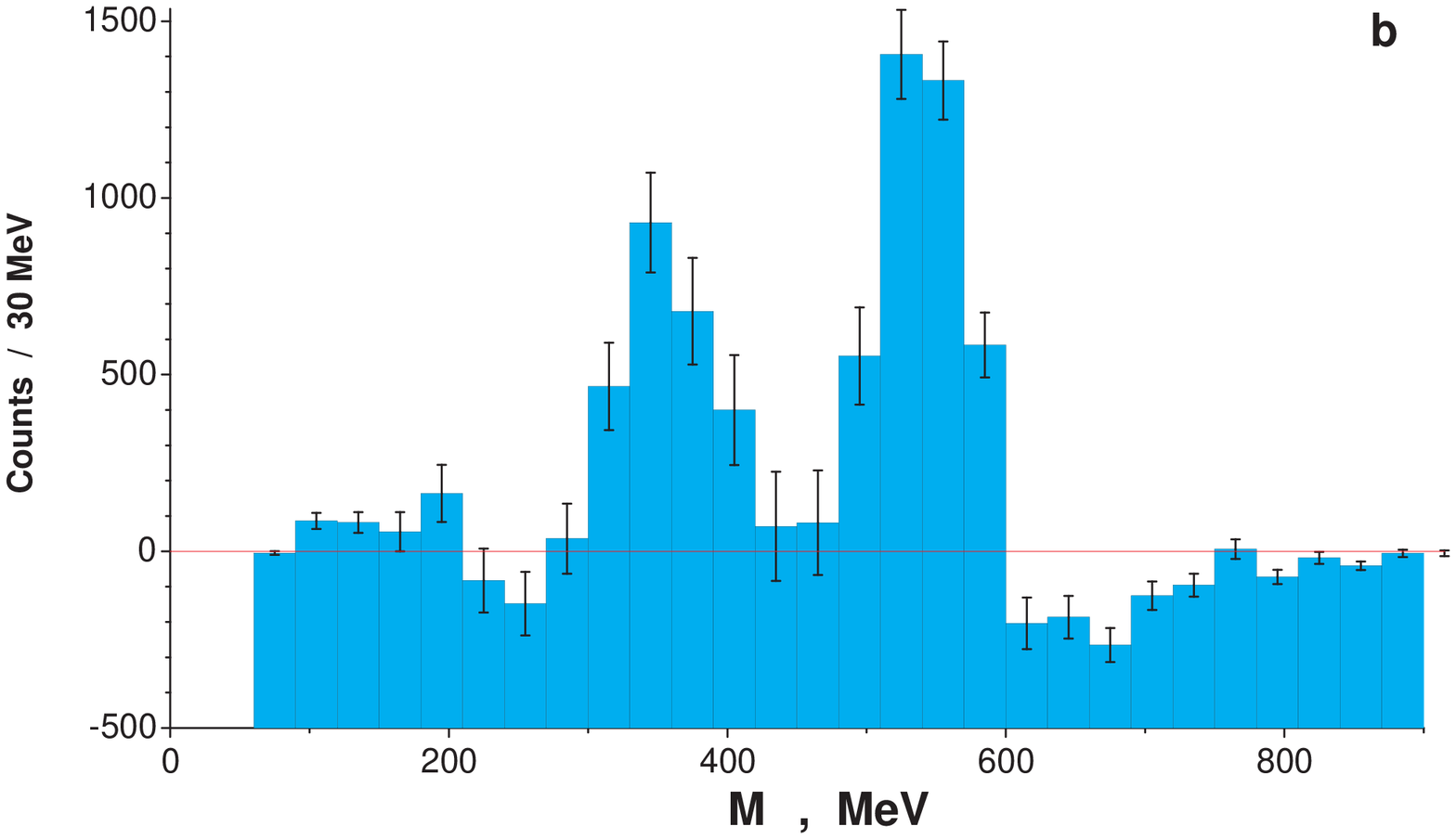}}
  \caption{Invariant mass distributions of pairs of $\gamma$-quanta in reaction (1)
 after background subtraction. In Fig. 2a, the background is normalized
to the total number of pairs in the spectrum; in Fig. 2b, with allowance made for the presence of
resonances in the spectrum (normalization coefficient is 0.976).
  }
\end{figure}

\section{Spectra of invariant masses of pairs of $\gamma$-quanta}

The invariant mass distribution of pairs of $\gamma$-quanta
($\gamma$-quanta from different arms of the spectrometer) after background subtraction
is shown in Fig. 2.  For the combinatorial background suppression, the following
selection criteria were used:

a) the number of $\gamma$-quanta in an event, $N_\gamma =2$;

b) the energies of $\gamma$-quanta, $E_\gamma > 100$ MeV.

 The invariant mass distribution of combinations of two $\gamma$-quanta selected by
random sampling from different events (the so-called event mixing) were used to estimate the background.

The absence of the peak corresponding to $\pi^0$-meson is due to the criteria of selection of events:
the setup is triggered if two or more groups of modules with the total energy $> 300$ MeV are available
(see also Fig. 3);
therefore, $\pi^0$-mesons were mainly registered in events with $N_{\gamma} > 2$ (a minimal opening angle of
$\gamma$-quanta registered by the setup equals $42^\circ$, see [2,3]).

The total (without subtraction of the background) spectra of the invariant masses of pairs of
$\gamma$-quanta and energy distribution of $\gamma$-quanta
are shown in Fig. 3.

\begin{figure}[t]
% \hspace*{3mm}\includegraphics[width=16cm]{ris3aw1.eps}
  \hspace*{3mm}\includegraphics[width=16cm]{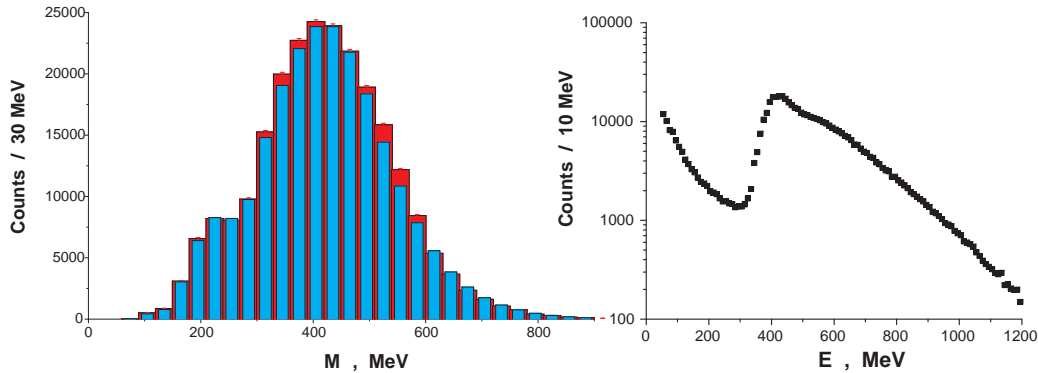}
  \caption{
Invariant mass distribution of pairs of $\gamma$-quanta and energy distribution of
$\gamma$-quanta in reaction (1) in selecting events with the number of
$\gamma$-quanta equal to 2.
  }
\end{figure}
Thus, in the reaction $d+C\to\gamma +\gamma +x$ at momentum 2.75 GeV/c per nucleon one can observe a
pronounced peak in the interval 300-420 MeV in the invariant mass spectrum of two
$\gamma$-quanta (see Fig. 2b). The parameters of the observed peak are given in more detail below
(see Figs. 4 and 5).

The errors displayed in Figs. 2,3 are statistical.

\section{Systematic errors}

Systematic errors may be due to:

    $\bullet$ errors in measurements of energies of $\gamma$-quanta,

    $\bullet$ errors in estimates of a combinatorial background.

The method of energy reconstruction of events is described in detail in [6,7]. One of the criteria of
accuracy of energy reconstruction is the conformity of the positions of peaks corresponding to the known
particles with their table mass values. As is seen in Fig. 2, the position of the peak corresponding to
$\eta$-mesons (as well as to $\pi^0$-mesons, see [2]) is in qualitative agreement
with the table values of their masses. A more precise determination of the position of peaks requires
minimization of systematic errors in describing the background which arise, in particular, due to the
violation of the energy-momentum conservation laws in selecting $\gamma$-quanta by random sampling from
different events (for instance, this is the reason for the presence of a traditional dip following the peak
of $\eta$-mesons).

\begin{figure}[t]
  \centerline{\includegraphics[width=10cm]{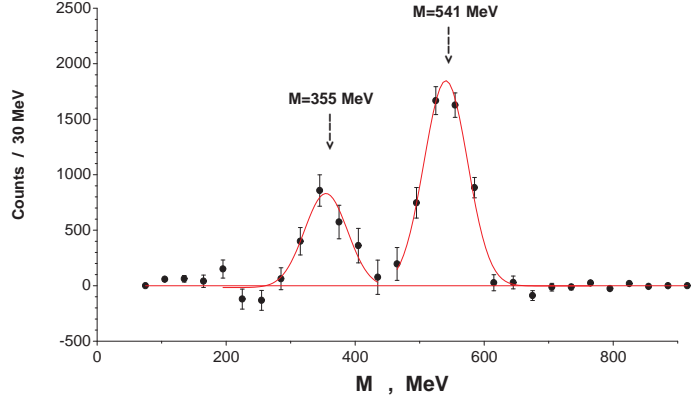}}
  \caption{
Invariant mass distribution of pairs of $\gamma$-quanta in reaction (1)
after the background subtraction by selecting accidentally $\gamma$-quanta
from different events under criteria  1) and 2) (see the text).
  }
\end{figure}
Figure 4 shows the distribution over the invariant mass
$M_{\gamma\gamma}$ of pairs of $\gamma$-quanta after the background subtraction. It is obtained by
random sampling of $\gamma$-quanta from different events under the following selection criteria:

1) the summed energy in an event $\le 1.7$ GeV (about 99\% of all events),

2) the sum of energies of $\gamma$-quanta selected accidentally from different events
$E_{\gamma 1} + E_{\gamma 2} \le 1.7$ GeV.
\begin{figure}[t]
  \centerline{\includegraphics[width=10cm]{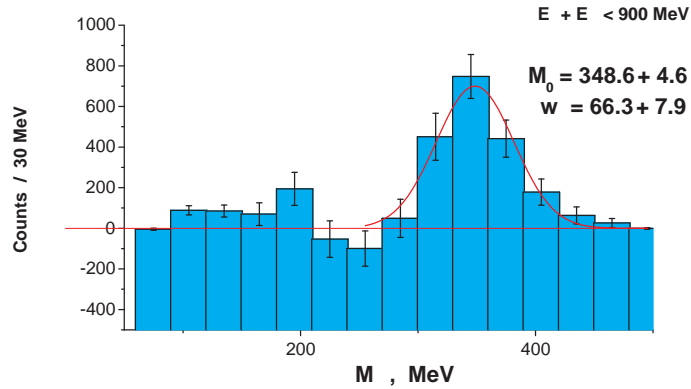}}
  \caption{
Invariant mass distribution of pairs of $\gamma$-quanta in reaction (1)
with the background subtracted in selecting pairs with the total energy less than 900 MeV.
  }
\end{figure}

The curves in Fig. 4 are the result of data approximation by the Gauss distribution,
including an additional free parameter $y_0$:
$$\frac {dN} {dM}=y_0+\frac {2N_0} {w\sqrt{2\pi}} e^{-\frac {2(M-M_0)^2} {w^2}}
\quad (1/MeV).\eqno(2)$$
The values for the parameters $y_0,~ N_0,~ w,~ M_0$ and $\chi^2$ are given below.

In the interval $0.18 \le M_{\gamma\gamma}\le 0.45$ GeV:

$M_0=354.6\pm 5.5$ MeV; $w=66.8\pm 12.0$ MeV; $N_0=2357\pm 453$; $y_0=-0.5\pm 1.7$; $\chi^2=1.8$.

In the interval $0.45 \le M_{\gamma\gamma}\le 0.78$ GeV:

$M_0=540.5\pm 2.1$ MeV; $w=67.2\pm 4.0$ MeV; $N_0=5199\pm 291$; $y_0=-0.08\pm 0.43$; $\chi^2=1.6$.

The signals-to-background ratios for the invariant mass intervals 300-420 MeV and 480-600 MeV (the vicinity
of $\eta$-meson mass) are $2.7\cdot 10^{-2}$ and $8.9\cdot 10^{-2}$, respectively. For comparison, analogous
values without the background suppression (without the selection criteria a) and b) of Section 3) are
$(4.0\pm 1.4)\cdot 10^{-3}$ and $3.2\cdot 10^{-2}$.

To determine the parameters of the observed peak more precisely, pairs of $\gamma$-quanta with the
total energy $E_{\gamma 1} + E_{\gamma 2} \le 0.9$ GeV (Fig. 5) were analyzed. Under the above
conditions the efficiency of recording $\eta$-mesons is almost equal to zero (a maximal opening angle of
$\gamma$-quanta registered by the setup equals $66^\circ$).

The curve in Fig. 5 is the result of data approximation by the Gauss distribution (2)
in the interval $0.24\div 0.51$ GeV. Below we give the values for the parameters
$y_0,~N_0,~w,~ M_0,$ and  $\chi^2$:

$M_0=348.6\pm 4.6$ MeV; $w=66.3\pm 7.9$ MeV; $N_0=1940\pm 228$; $y_0=0.003\pm 0.120$; $\chi^2=1.04$.

Thus, the position and width of the peak corresponding to $\eta$-meson (see Fig. 4) are in good agreement
with the table value of its mass (systematic errors do not exceed 1.5\%) and with the mass resolution
of the instrumentation (see eq. (7) below). The total number of events registered
in the $\eta$-meson region 450-660 MeV after background subtraction is $5177 \pm 293$.

The width estimates of the observed new resonance structure are given in Section 6.

\vspace*{3mm}

\section{Analysis of the obtained data}

To elucidate the nature of the observed peak, we have investigated the dependence of
its position and width on

 $\bullet$ an interval of the opening angles of $\gamma$-quanta,

 $\bullet$ a level of energy selection of $\gamma$-quanta,

 $\bullet$ a level of summed energy selection of pairs of $\gamma$-quanta. \\
Figure 6 shows the invariant mass distribution of pairs of $\gamma$-quanta for two different
intervals of the opening angle of $\gamma$-quanta.

\begin{figure}[h]
 \centerline{\includegraphics[width=8cm]{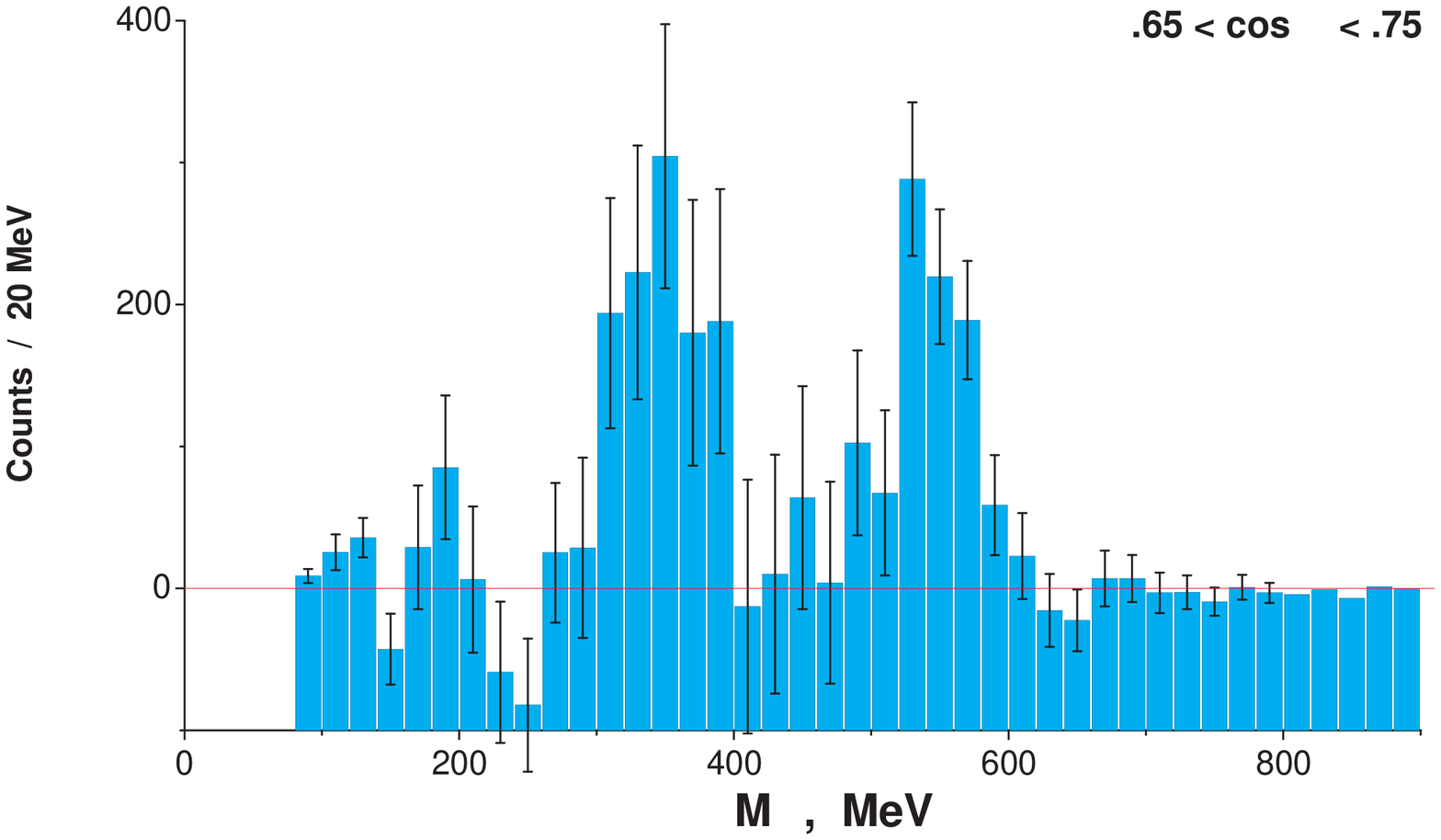}
  \includegraphics[width=8cm]{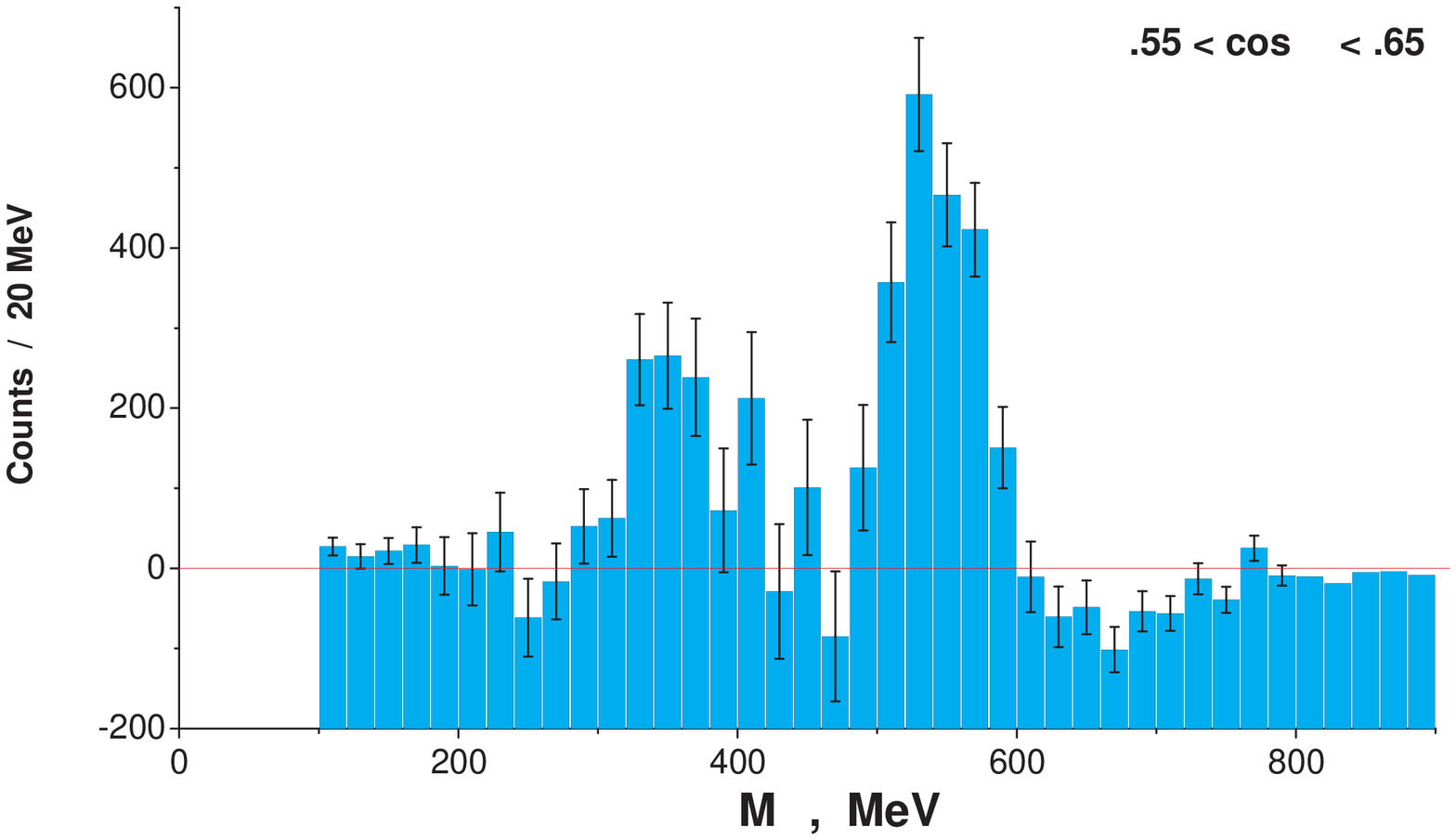}}
  \caption{
The same as in Fig. 2b, for two different intervals of the opening angle of $\gamma$-quanta.
  }
\end{figure}

\begin{figure}[h]
 \centerline{
 \includegraphics[width=8cm]{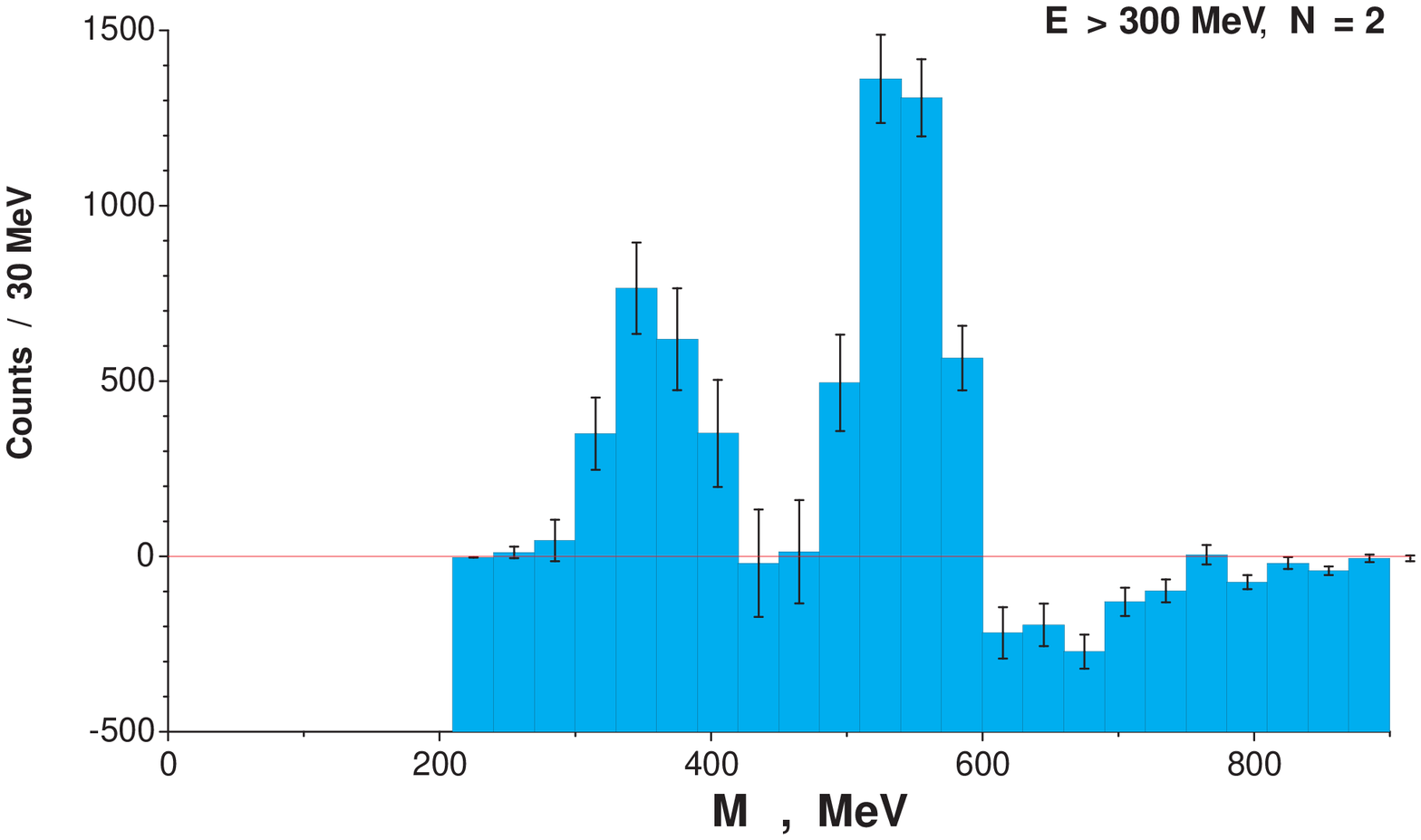}
 \includegraphics[width=8cm]{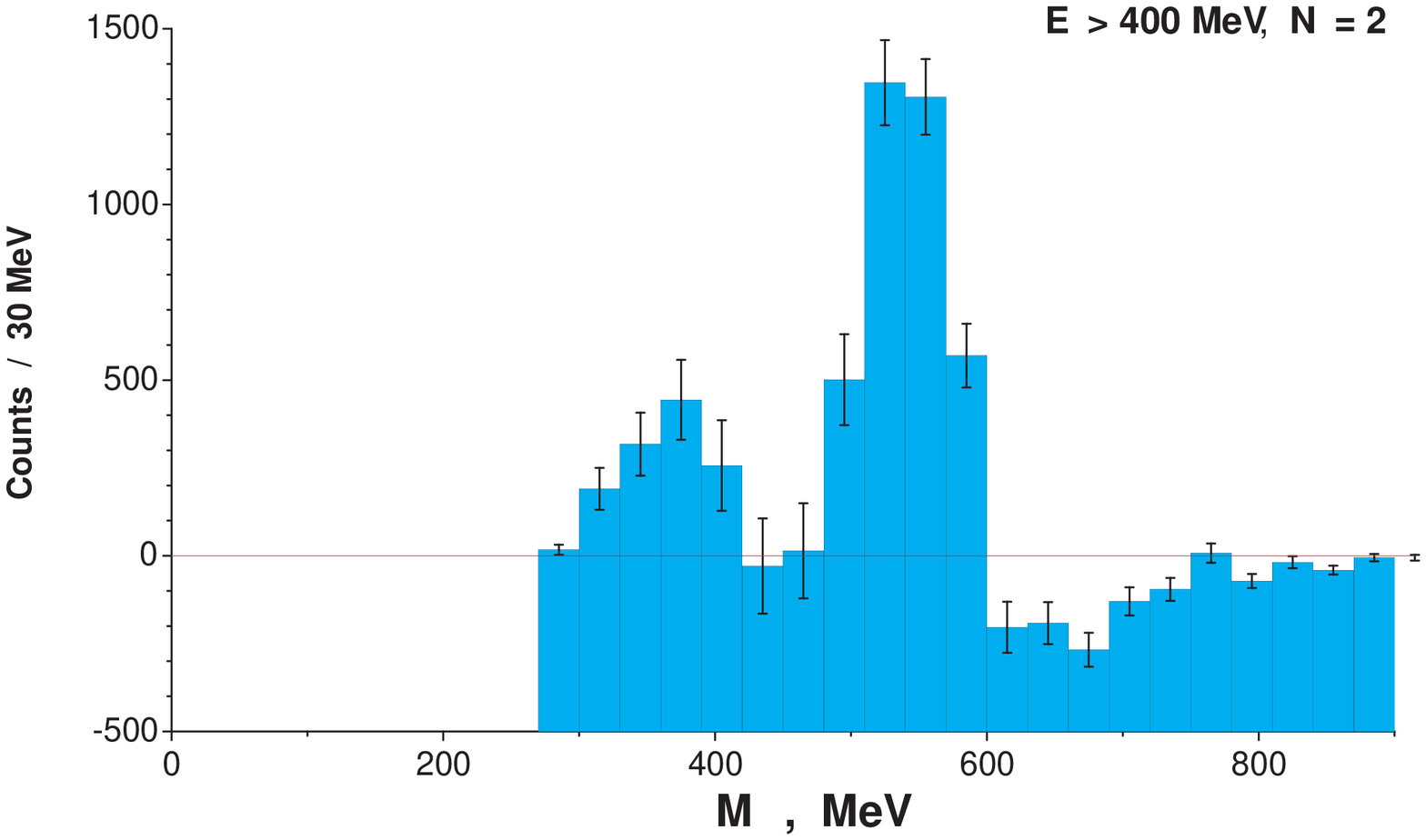}}
  \caption{
 The same as in Fig. 2b, at different levels of energy selection of $\gamma$-quanta.
 }
\end{figure}

The invariant mass spectra at different levels of energy selection of $\gamma$-quanta are displayed
in Fig. 7.

The invariant mass spectra in different intervals of summed energy of two $\gamma$-quanta are shown
in Fig. 5 ($E_1 + E_2 < 900$ MeV) and Fig. 8 ($E_1 + E_2 > 900$ MeV).

\begin{figure}[h]
 %\hspace*{3cm}\includegraphics[width=10cm]{ris8aw1.eps}
  \hspace*{3cm}\includegraphics[width=10cm]{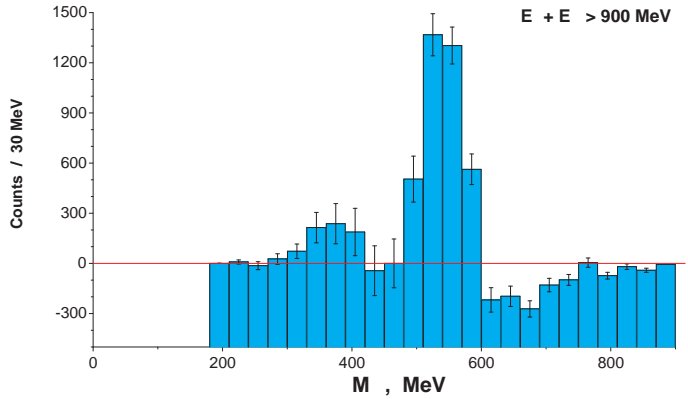}
  \caption{
 The same as in Fig. 2b, at the total energy of two $\gamma$-quanta, $E_1 + E_2 > 900$ MeV.
 }
\end{figure}

As is seen from Figs. 5-8, the position and width of the observed peak remain almost unchanged in
different intervals of both the energies and the opening angles of $\gamma$-quanta:
the mean value of the mass under different conditions varies in the range
$348\div 365$ MeV. The total number of events registered in the region 280-450 MeV (a summed number
of pairs in the histograms in Figs. 5 and 8) is $2680\pm 310$.
\vspace*{3mm}

\section{Estimates of the width and production cross section of the observed
resonance structure}

Values for the width $w$ in (2) are also specified by the instrumentation resolution.
The corresponding contribution to $w$ can be estimated by the formula
$$(1/4)w_{app}^2=(\frac {\partial M } {\partial E_1} \Delta E_1)^2+
(\frac {\partial M } {\partial E_2} \Delta E_2)^2+
(\frac {\partial M } {\partial \theta_{\gamma\gamma}} \Delta \theta_{\gamma\gamma})^2 , \eqno(3)$$
where $\Delta E_1,~ \Delta E_2$, and $\Delta \theta_{\gamma\gamma}$ are standard errors in measurements
of $\gamma$-quanta energies and the opening angle,
$$\Delta E_i\simeq 0.068\cdot\sqrt{E_i},~ E_i(GeV),~i=1,2;$$
$$\Delta \theta_{\gamma\gamma}\le \sqrt{2}\cdot \frac {12cm} {\sqrt{12}}\cdot
\frac {1} {300cm}=0.016.$$

Energy distributions of $\gamma$-quanta in combinations with the invariant mass $340\le M \le 360$ MeV
are displayed in Fig. 9.

\begin{figure}[t]
\centerline{ \includegraphics[width=7.5cm]{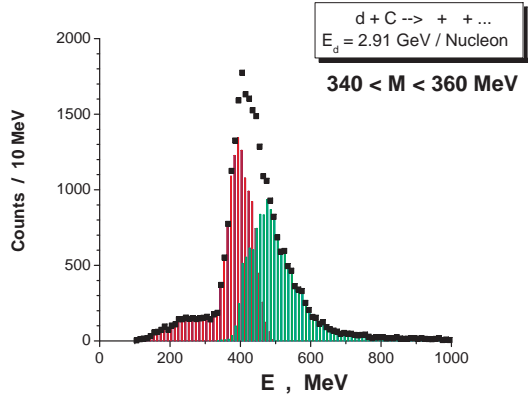}}
  \caption{
 Energy distributions of $\gamma$-quanta in combinations with the invariant mass in the interval
340-360 MeV.
The histograms show the distributions of $\gamma$-quanta with the lowest and highest energy in each pair.
 }
\end{figure}

Mean energy values for $\gamma$-quanta in a pair in the given interval of invariant masses amounts to
$E_1\simeq 380$ MeV; $E_2\simeq 520$ MeV.

Analogous values for the invariant mass interval 540-560 MeV (in the vicinity of the $\eta$-meson mass)
are $E_1\simeq 520$ MeV; $E_2\simeq 750$ MeV.

Below we give the results of calculations by formula (3) with the use of the above values for $E_1$
and $E_2$:
$$w_{app} (340<M<360 MeV)=52.6~MeV, \eqno(4)$$
$$w_{app} (540<M<560 MeV)=68.6~MeV. \eqno(5)$$
Correspondingly, the intrinsic widths of detected resonances are
$$w_{intr} (R\to\gamma\gamma)\simeq (w^2-w_{app}^2)^{1/2}\simeq 40.6\pm 11.8,\eqno(6)$$
$$w_{intr} (\eta\to\gamma\gamma)\simeq (w^2-w_{app}^2)^{1/2}\approx 0.\eqno(7)$$

As one can see, a width of the $\eta$-meson, as expected, is practically equal to 0, whereas it essentially
differs from zero in the observed resonance. The value of the width $\Gamma$ in the Breit-Wigner function (see Fig. 10)
practically coincides with $w$ in Gauss distribution (2); thus, the intrinsic width of the
observed resonance structure is about $41\pm 12$ MeV.
\begin{figure}
\centerline{ \includegraphics[width=10cm]{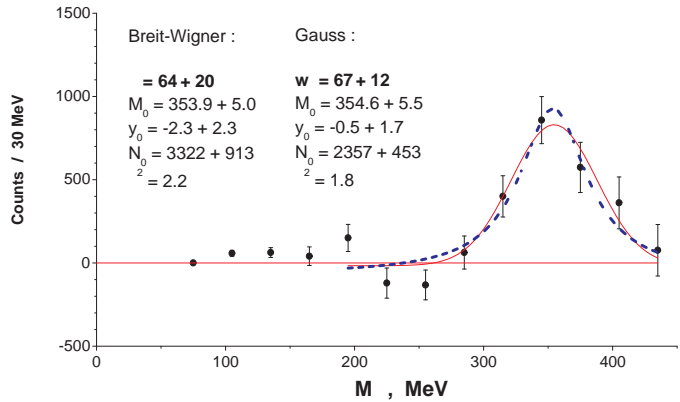}}
  \caption{
 Comparison of the data (see fig. 4) approximation by the Breit-Wigner function
$\frac {dN} {dM}=y_0+(2N_0 / \pi )\cdot [\Gamma/(4(M-M_0)^2 + \Gamma^2)]$ (the dashed line)
and by Gauss distribution.
 }
\end{figure}

It should be emphasized that the values for both the mass and width of the observed
resonance structure have to
be determined more accurately. Some shift of these values may be due to strict selection criteria of
events in energy which are more appropriate for registration of $\eta$-mesons.

%\vspace*{3mm}

%\section{Estimate of production cross section}

The summed number of $dC$-interactions in the experiment amounts  to $\sim 3\cdot 10^{12}$. With
allowance made for the efficiency of registration, $\varepsilon\sim 10^{-3}$, the cross section of
the observed process (for the total cross section of $dC$-interactions the value 612 mb was used [8]) is
$$\sigma_{\gamma\gamma}\approx \frac {2.7\cdot 10^3} {3\cdot 10^{12}\cdot\varepsilon}
\cdot\sigma_{tot} (dC)\sim 0.6~ \mu b .$$

\vspace*{3mm}

\section{Conclusion}

Thus, based on a thorough analysis of experimental data measured at the JINR Nuclotron and record
statistics of $2680 \pm 310$ events of $1.5\cdot 10^6$ triggered interactions of a total number
$3\cdot 10^{12}$ dC-interactions
there was observed a new resonance structure with the mass $M= 355\pm 6\pm 9$ MeV, width $\Gamma=41\pm 12$ MeV,
and preliminary production cross section $\sigma_{\gamma\gamma}\sim 0.6~ \mu b$
in the invariant mass spectrum of two $\gamma$-quanta produced
in $dC$-interactions at momentum of incident deuterons 2.75 GeV/c per nucleon. To verify the above
conclusions, measurement of quantum numbers, and more accurate determination of mass, width and
cross section of the observed resonance structure, new experiments are required to be carried out under
conditions appropriate for registration of pairs of $\gamma$-quanta with the invariant mass 300-400 MeV.
We hope to conduct detailed analysis of the available experimental data, and dwell upon a physical
character and possible interpretations of the observed resonance in our further publications.

\vspace*{8mm}
 {\bf Acknowledgements}

We are grateful to S.B. Gerasimov, M.I. Gorenstein, G.M. Zinovjev, M.K. Suleymanov,
O.V. Teryaev, V.D. Toneev, A.S. Khrykin, and V.L. Yudichev for numerous fruitful discussions.
We thank V.V. Arkhipov, A.F. Elishev, A.D. Kovalenko, A.I. Malakhov, G.L. Melkumov,
S.G. Reznikov, and the staff of the Nuclotron for their help in conducting the experiment,
as well as
A.V. Belozerov, M.A. Kozhin, M.A. Nazarenko, S.A. Nikitin and V.V.Skokov for their help in analyzing data.
We would like also to express special thanks to V.G. Kadyshevsky, V.A. Matveev and A.N. Tavkhelidze
for their interest to this paper.

The work was supported in part by the Russian Foundation for Basic Research, grant
05-02-17695 and a special program of the Ministry of Education and Science of the Russian Federation,
grant RNP.2.1.1.5409.

\end{document}